\documentclass[twocolumn,tighten]{aastex63}
\usepackage{tabularx}
\usepackage{url}
\usepackage{graphicx}
\usepackage[T1]{fontenc}
\usepackage{ae,aecompl}
\usepackage{booktabs}

\shortauthors{Bhattacharyya et al.}

\graphicspath{{./}{figures/}}

\def\ltsim{\mathrel{\hbox{\rlap{\hbox{\lower3pt\hbox{$\sim$}}}\hbox{\raise2pt\hbox{$<$}}}}}
\def\gtrsim{\mathrel{\hbox{\rlap{\hbox{\lower3pt\hbox{$\sim$}}}\hbox{\raise2pt\hbox{$>$}}}}}

\newcommand\asca{{\it ASCA}}
\newcommand\swift{{\it Swift}}
\newcommand\xmm{{\it XMM-Newton}}
\newcommand\astst{{\it AstroSat}}

\begin{document}

\title{Blazar Variability: A Study of Non-stationarity and the Flux-RMS Relation}

\correspondingauthor{Souradip Bhattacharyya}
\email{sigmabeta896@gmail.com}

\author{Souradip Bhattacharyya}
\affiliation{Department of Physics, Presidency University, 86/1 College Street, Kolkata, West Bengal, India 700073.}

\author{Ritesh Ghosh}
\affiliation{Visva-Bharati University, Santiniketan, Bolpur, West Bengal, India 731235.}

\author{Ritaban Chatterjee}
\affiliation{Department of Physics, Presidency University, 86/1 College Street, Kolkata, West Bengal, India 700073.}

\author{Nabanita Das}
\affiliation{Department of Physics, Presidency University, 86/1 College Street, Kolkata, West Bengal, India 700073.}


\begin{abstract}
We analyze X-ray light curves of the blazars Mrk~421, PKS~2155-304, and 3C~273 using observations by the Soft X-ray Telescope on board \astst{} and archival \xmm{} data. We use light curves of length 30-90 ks each from 3-4 epochs for all three blazars. We apply the autoregressive integrated moving average (ARIMA) model which indicates the variability is consistent with short memory processes for most of the epochs. We show that the power spectral density (PSD) of the X-ray variability of the individual blazars are consistent within uncertainties across the epochs. This implies that the construction of broadband PSD using light curves from different epochs is accurate. However, using certain properties of the variance of the light curves and its segments, we show that the blazars exhibit hints of non-stationarity beyond that due to their characteristic red noise nature in some of those observations. We find a linear relationship between the root-mean-squared amplitude of variability at shorter timescales and the mean flux level at longer timescales for light curves of Mrk~421 across epochs separated by decades as well as light curves spanning 5 days and $\sim$10 yr. The presence of flux-rms relation over very different timescales may imply that, similar to the X-ray binaries and Seyfert galaxies, longer and shorter timescale variability are connected in blazars. 
\end{abstract}


\section{Introduction} \label{sec:intro}
Variability of emission can provide crucial information about the physics of accretion and outflow in active galactic nuclei (AGN). A common tool that is used to probe the variability properties of AGN in the spectral domain is the power spectral density (PSD), which is a measure of the amplitude of variability across different Fourier frequencies. Alongwith this there exist endeavours to model the fluctuations as stochastic variability in the time domain. Regardless of the approach, there is a need to identify the statistical nature of the variability, with stationarity being one of the important characteristics. In the strictest sense, stationarity implies that the mean, variance and all higher order moments associated with the process remain constant throughout. However this represents an idealization, and the definition is relaxed for an actual time series for what is called a weakly stationary process, in which the mean and second order statistics like the autocorrelation function (ACF) and PSD remain constant.

The nature of AGN variability, particularly stationarity, depends on the order of the timescale that is being probed \citep{2020arXiv200108314S}. The X-ray PSD of some AGN may be modeled as a piece-wise power law, the slope of which is $\sim -1.0$ at longer and $\sim -2.5$ at shorter timescales with a characteristic timescale $T_B$ ($\sim$days) at which the slope changes \citep{2002MNRAS.332..231U, 2009ApJ...704.1689C, 2011ApJ...734...43C}. Therefore, at shorter timescales ($\sim$day) the variability has a red noise nature, a process which is non-stationary. The red noise nature may extend to $\sim$months to $\sim$yr in other AGN, in which the value of $T_B$ is larger. For a red noise process, the mean and variance computed from segments of a light curve fluctuate with time irrespective of the duration of the segements \citep[e.g.,][]{2003MNRAS.345.1271V}. Realistically, at longer timescales ($\gg$ yr), the slope of the PSD tends to zero, and the mean and variance have to converge to constant values \citep{2005MNRAS.359..345U}. The latter feature is known as asymptotic stationarity. Between these two timescales the assumption of weak stationarity may hold. On the other hand, if the PSD of the underlying process or some charateristic timescale dependent on the physical conditions in the AGN changes then \textit{local} non-stationarity- a form of variability different from the \textit{global} non-stationarity due to red noise variability would arise \citep{10.1144/IAVCEI001.11}. One of our primary goals in this paper is to identify local non-stationarity in blazar light curves the reason for which is described below.

In black hole X-ray binaries (BHXRBs), PSD of variability has been observed to undergo transition between low and high spectral states taking place at $\sim$months to $\sim$yr timescale as well as minor changes in the PSD amplitude and slope over timescales of days to weeks \citep{2006ARA&A..44...49R}. However, if it is assumed that the accretion processes in BHXRBs and AGN are similar and the differences in the characteristic length and timescales are scaled only by black hole mass \citep{2001MNRAS.323L..26U, 2006Natur.444..730M} then changes in the shape of the PSD of AGN variability due to transition between spectral states may be observed at timescales of centuries or more.
Meanwhile, evidence of local non-stationarity in AGN variability at timescales of $\lesssim$ day, has been found in the Narrow Line Seyfert-1 galaxy IRAS 13224-3809 \citep{2019MNRAS.482.2088A}, which exhibits changes in the shape and amplitude of its time resolved PSD. 

Blazars are a class of active galactic nuclei with a prominent jet within a few degrees of our line of sight \citep{1995PASP..107..803U}. Due to relativistic beaming apparent emission from the jet is amplified by an order of magnitude or more in the observer's frame and hence the total X-ray emission from blazars is dominated by that from their jets.  Recently, the X-ray PSD of a blazar have been confirmed to be a power-law with a break \citep{2018ApJ...859L..21C}. This property is similar to those of some other classes of accreting black hole systems, e.g., Seyfert galaxies and some BHXRBs, in which the X-ray variability is primarily from the fluctuations in the accretion disk region \citep{1997MNRAS.292..679L} and the characteristic break timescale is proportional to the black hole mass \citep{2002MNRAS.332..231U,2006Natur.444..730M,2009ApJ...704.1689C,2011ApJ...734...43C}. These fluctuations can influence the X-ray variability from the jet if a connection between the disk and the jet exists. Thereby the characteristic timescale is imprinted in the jet variability, which manifests as a break in the blazar X-ray PSD, which can provide crucial constraints on the theory of jet launching and collimation \citep[e.g.,][]{2018MNRAS.480.2054M,2018ApJ...859L..21C}. 

However, to confirm the existence of such a break or more generally to constrain the shape and amplitude of the broadband PSD of blazars it needs to be tested if short-timescale ($\sim$minutes to $\sim$days) X-ray PSD of blazars remain the same in shape and amplitude such that a broadband PSD may be constructed accurately using data from multiple epochs. So far rigorous studies of such local non-stationarity in blazar variability have been scant. One of the first significant demonstrations have been with the \xmm{} EPIC-pn light curves of the BL Lac object PKS~2155-304 \citep{2005ApJ...629..686Z} especially during the flaring states, in which it exhibits significant changes in the excess variance and fractional rms variability amplitude from one time interval to another separated by a few hours. 

In this paper, we start by rigorously establishing the global non-stationarity in the X-ray variability of blazars that is expected from their red noise nature. We then attempt to fit auto-regressive integrated moving average (ARIMA) models  \citep{1981ApJS...45....1S, 2018FrP.....6...80F} to the data and compare the best-fit models across the epochs. This is one of the first such attempts directed at sub-day variability of blazars, which may provide quantitative information regarding the extent of the non-stationarity of their fluctuations. 

Moving on to the search for signs of local non-stationarity, we compute the X-ray PSD of three blazars using well-sampled light curves from different epochs to test whether there is a significant change in their shapes. Due to the inherent biases introduced by red noise processes \citep{1993MNRAS.261..612P}, it is important to determine the PSD and its uncertainties rigorously from the available data so that the comparison between the epochs is accurate. For that purpose, we apply a Monte-Carlo type technique to the X-ray light curves of the three blazars obtained from \astst{}  and \xmm{} each of which spans $30-90$ ks and are separated by a few years. \citet{2003MNRAS.345.1271V} provided a framework for diagnosing local non-stationarity in a light curve- namely the scatter of excess variance and fractional variance, which we apply to our sample of observations.

Furthermore, the flux variability of accreting compact objects, including AGN and BHXRBs, have been shown to exhibit the so-called ``flux-rms relation,'' i.e., a linear relationship between the root-mean-squared amplitude of variability at shorter timescale and the mean flux level at longer timescale \citep{2001MNRAS.323L..26U,2004MNRAS.348..783M,Gaskell_2004,2005MNRAS.363..586U,2011MNRAS.413.2489V}. The flux-rms relation is exhibited by BHXRBs at different flux states characterized by different shapes of the PSD indicating that the former is a more universal property of the variability than the PSD shape \citep{2004A&A...414.1091G,2012MNRAS.422.2620H}. \citet{2005MNRAS.359..345U} showed that the flux-rms relation mathematically implies that the variability is non-linear, which in turn indicates that the distribution of the fluxes is log-normal although these have been questioned recently by \citet{2020arXiv200108314S}. 

The flux-rms relation implies that the variability at shorter timescales is related to or coupled with that at longer timescales. This puts strong constraints on the models of flux variability of accreting compact objects. For example, theories similar to the propagating fluctuation model  \citep{1997MNRAS.292..679L}, in which the longer-term fluctuations from the outer disk propagate inward along with the accretion flow and modulate the shorter timescale variability generated at the smaller radii, are favored. On the other hand, blazar variability is usually assumed to be due to a combination of several processes, namely, acceleration of particles by shocks passing through the jet or magnetic reconnection \citep[e.g.,][]{1985ApJ...298..114M,2014ApJ...785..132J,2011ApJ...727...21J,2015ApJ...815..101N}, cooling of those particles though radiation and adiabatic expansion, and turbulence in the magnetic field and density \citep{Marscher_2013}. It is theoretically unclear if such a variability should exhibit the flux-rms relation. The first evidence of a linear flux-rms relation and log-normal flux distribution in a blazar were found in the RXTE-PCA data of BL Lacertae \citep{2009A&A...503..797G}. Similar signs were found in the Kepler light curves of the BL Lac object W2R1926+42 \citep{2013ApJ...766...16E} as well as in the Fermi-LAT light curves of multiple blazars \citep{2017ApJ...849..138K}. The log-normal distribution of flux has been found to be present in multiwavelength observations of PKS~2155-304 and Mrk~421 \citep{Chevalier:2015gia,2016A&A...591A..83S,2020ApJ...891..120B}. In this paper, we analyze the X-ray light curves of Mrk~421 at multiple epochs to search for a relation between the rms and flux at a longer range of timescales. 

In {\S}\ref{sec:datared} we describe the reduction of data from \xmm{} and \astst{} , we carry out the analyses related to ARIMA, PSD, variance, and flux-rms relation in {\S}\ref{sec:temporalanalysis}, {\S}\ref{sec:spectralanalysis},{\S}\ref{sec:variance} and {\S}\ref{sec:linearfluxrms}, respectively and finally we summarize the results and discuss the implications in {\S}\ref{sec:disucss}.

\section{observations and Data Reduction} \label{sec:datared}
We search NASA's High Energy Astrophysics Science Archive Research Center (HEASARC) to find X-ray light curves of Fermi-detected blazars \citep{2010ApJ...719.1433A}. Our selection criteria were to find blazars that have multiple light curves of length longer than $30$ ks and separated by at least a year. The signal-to-noise ratio (SNR), defined as the mean flux divided by the mean flux uncertainty, is used to reject light curves with SNR<10. We list the observations of the sources in Table \ref{Table:obs}. We use a total of 9 sub-day scale observations of Mrk~421, 3C~273 and PKS~2155-304 from \xmm{}, 3 observations of Mrk~421 from \astst{}  \citep{2016arXiv160806051R} and \asca{} pertaining to timescales of a few days, and a decade-long monitoring light curve of Mrk~421 from \swift{}.
 
\subsection{\xmm{} observations}
The \xmm{} data sets are reduced using the scientific analysis system (SAS) software (version 16.1.0) and the latest calibration database available. The task {\it epchain} is used to produce the calibrated event files of the EPIC-pn \citep{2001A&A...375L...5S} data because of its higher signal to noise ratio as compared to MOS. We use the single and double pixel events with pattern 0--4 only to filter the processed \textit{pn} event file. We create a GTI (Good Time Interval) file, using the task {\it tabgtigen}, to correct for particle background counts. We use a rate cutoff of $< 1 \, \rm ct\,s^{-1}$ for energy $>10$ keV.  
We use a circle of radius $40"$ to extract the source region. The centroid of the source is chosen as the center of the circle. We select a circle of radius $40"$ located on the same CCD but in a source free area as the background region. We use the command {\it epatplot} in SAS to check for possible pile up in the source. We find that none of the sources contains any pile up. The source and background light curves are extracted from the cleaned PN event file using xselect V.2.4 d. The ftool task {\t lcmath} is used to create the background subtracted light curves.  

\subsection{\astst{} observations} 
\begin{figure*}
\begin{center}
\includegraphics[scale=0.79,trim={0.5cm 7cm 0.6cm 7cm},clip]{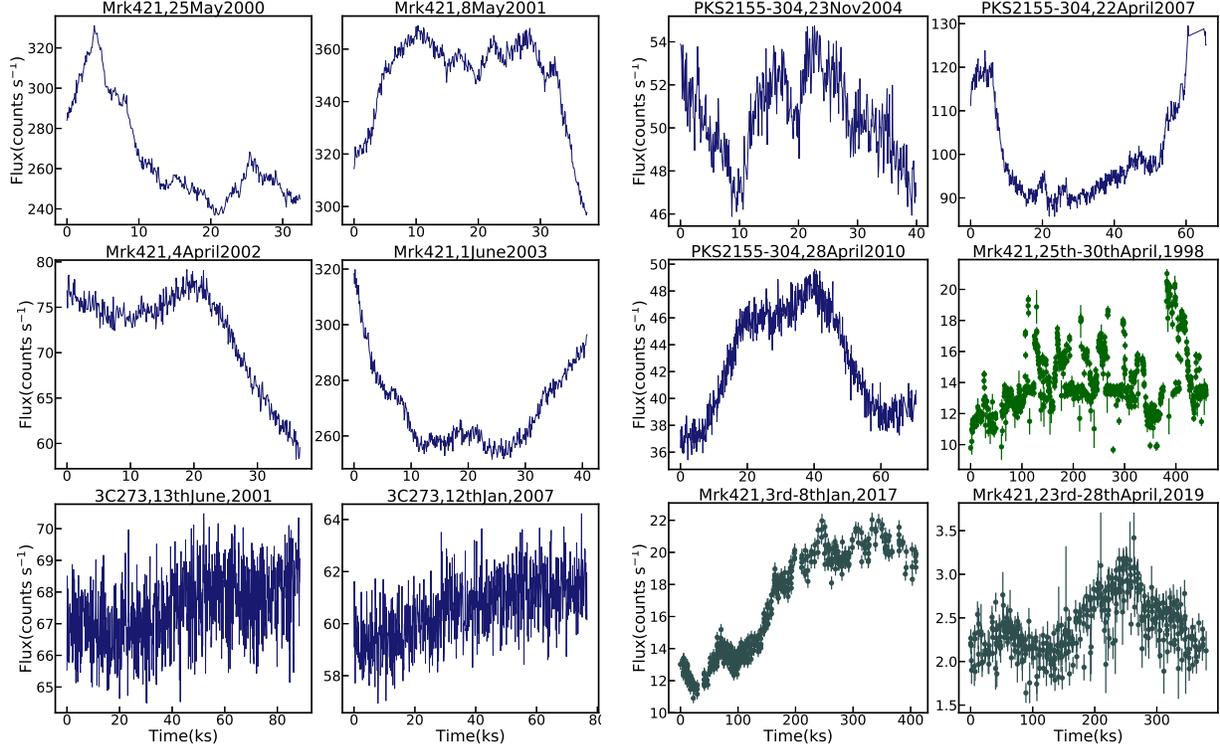}
\caption{X-ray light curves from sub-day observations of Mrk~421, 3C~273, PKS~2155-304 and sub-week observations of Mrk~421 using \xmm{} (\textit{blue}), \asca{} (\textit{green}) and \astst{} (\textit{gray}). \label{Figure:lc}}
\end{center}
\end{figure*}

The observations of Mrk~421 were carried out from 2019 April 23--28 using the Soft X-ray Telescope (SXT) instrument \citep{2017JApA...38...29S}. The data are collected in the photon count (pc) mode. We process the raw data and generate level-2 products using {\it sxtpipeline} software, provided in the SXT data analysis package (AS1SXTLEVEL12-1.4b)\footnote{\url{http://www.tifr.res.in/~astrosat_sxt/sxtpipeline.html}}. The pipeline tool creates orbit-wise level-2 clean event files, which are then combined using another tool {\it sxtevtmerger}. For further analysis, we use various tools of {\it heasoft} (v. 6.25). Using {\it xselect}, we extract and plot the image on {\it ds9}. A circle of radius 1.511 arcmin is used to make the source region. To get the background flux, we construct multiple background regions centered at different points on the FOV with radii 2.523 arcmin, 2.188 arcmin and 1.669 arcmin each. The source and background filters are individually applied on the image and light curves for each are generated. We subtract the background noise from source using {\it lcmath}. The final light curve is produced using the {\it lcurve} tool with a binning interval of 300 s. All light curves are displayed in Figure \ref{Figure:lc}.

\begin{table*}
{\footnotesize
\centering
  \caption{Parameters of the X-ray light curves of the three blazars. \label{Table:obs}}
  \begin{tabular}{cccccccc} \hline\hline 

		Source &Instrument &Date &Referred &Duration, $T$ &N &Mean, $\bar{X}$ &SNR,  \\ 
			&	&   &as   &(ks) &  &(cts s$^{-1}$)  &$\bar{X}/\overline{\sigma_{err}}$ \\ \hline \\
		
		PKS~2155-304	&\xmm{} EPIC-pn &2004-11-23 &obs1-2004 &41 &400  &50.5 &56.1 \\
			& &2007-04-22		&obs2-2007 &65  &591  &98.6 &78.0	\\
			&	&2010-04-28	&obs3-2010	&65	 &686  &42.9 &51.4	 \\
		
		3C~273 &\xmm{} EPIC-pn  &2001-06-13 &obs1-2001 &90 &886  &67.4 &64.8	\\ 
			&	&2007-01-12 &obs2-2007 &79 &767  &60.5  &61.3	\\
		
		Mrk~421 &\xmm{} EPIC-pn  &2000-05-25 &obs1-2000 &33 &310  &267 &124	\\ 
			& &2001-05-8 &obs2-2001 &38 &374  &349  &143\\
		   & &2002-05-4 &obs3-2002 &37 &368  &72.6 &77.4	\\
			& &2003-06-1 &obs4-2003 &41 &407  &270 &122	\\
\hline
        Source &Instrument &Start &End &Duration  &N &$\bar{X}$ &SNR \\ 
			&	&   &   &(ks)  &  &(cts s$^{-1}$)  & \\ \hline \\
		
        Mrk~421 &\asca{} GIS-3 &1998-06-30 &1998-06-25 &459 &977  &13.9  &50.7 \\
        
        	&\swift -XRT &2007-10-7  &2017-06-26  &3.07$\times 10^5$  &765  &25.7 &137.8 \\
        
        	&\astst{} SXT &2017-01-3  &2017-01-8 &409  &495  &16.0  &43.0 \\
        
        	&\astst{} SXT	&2019-04-23  &2019-04-28  &379  &424  &2.41  &17.9 \\

\hline \hline
\end{tabular}  
}
\end{table*}

\section{Time Domain Analysis} \label{sec:temporalanalysis}

Light curves of AGNs, including some blazars, have been modeled as a parameterized stochastic process- either using the formalism of a mixed Ohrnstein-Uhlenbeck process \citep{2011ApJ...730...52K} or a continuous autoregressive moving average (CARMA) process \citep[e.g.,][]{2014ApJ...788...33K, 2019ApJ...885...12R}. The central assumption behind this is that the light curve is weakly stationary and is justifiably so given the long duration observations, e.g., of several years, to which these are applied. However, a light curve which is stationary at one timescale may be non-stationary at another. Such is expected from a process which has a broken power-law form of PSD, where global non-stationarity is expected in the high frequency part corresponding to shorter timescales. 
Here we perform a rigorous statistical analysis of the light curves spanning tens of ks to detect the violation of strict-sense stationarity. 

\subsection{Test of Non-stationarity}

The Augmented Dickey-Fuller (ADF) test \citep{10.1093/biomet/71.3.599} is widely used to test non-stationarity of time series. Here the null hypothesis is that a unit-root characteristic of a non-stationary process exists. Roughly, the more negative the test statistic is, it is more likely that the time series is stationary. We apply the ADF test on the time series $X_i$ and its first difference $Y_i$. The latter is obtained by differencing the light curve $X_i$ once. 

\begin{equation} \label{eq:diff}
Y_i= \nabla X_i = X_i-X_{i-1} = (1 - L) X_i,
\end{equation}

where $L$ is the lag operator. Should $Y_i$ be stationary while $X_i$ is not, this paves the way for application of autoregressive integrated moving average (ARIMA) model on $X_i$. Table \ref{Table:adf} displays the results of applying the ADF test on the \xmm{} light curves and their first differences. It shows that the null hypothesis cannot be rejected for any of the light curves but can be rejected for their first differences, i.e., all of those light curves exhibit non-stationarity but that can be removed by differencing once. The test qualifies the first differences as stationary with a p-value $< 0.01$, implying that ARIMA modeling may be applied to the \xmm{} light curves. In addition this application is possible because those light curves are very regularly sampled.

\begin{table}
\centering
{\footnotesize
  \caption{Results of the ADF test applied to the blazar light curves.  \label{Table:adf}}
  \begin{tabular}{cccc|cc} \hline\hline  
        \multicolumn{2}{c}{} & \multicolumn{2}{c|}{Actual light curve} &
\multicolumn{2}{c}{After differencing once} \\ \hline 
		Source &Observation &Test &p-value &Test &p-value \\
         & &statistic & &statistic &  \\ \hline
		PKS &obs1-2004 &$-1.27$ &$0.64$ &$-11.45$ &$<0.01$ \\
			2155-304&obs2-2007 &$-0.92$ &$0.78$ &$-3.97$ &$<0.01$ \\
			&obs3-2010	&$-1.42$ &$0.57$ &$-15.68$ &$<0.01$ \\
		3C~273 &obs1-2001 &$-1.93$ &$0.32$ &$-11.88$ &$<0.01$  \\
		    &obs2-2007	&$-1.69$ &$0.47$ &$-13.60$ &$<0.01$ \\
		Mrk~421  &obs1-2000 &$-1.64$ &$0.46$ &$-6.50$ &$<0.01$ \\ 
			&obs2-2001 &$-0.57$ &$0.88$ &$-3.40$ &$<0.02$ \\
		   &obs3-2002 &$2.09$ &$0.99$ &$-3.81$ &$<0.01$	\\
		   &obs4-2003 &$-0.31$ &$0.92$ &$-3.50$ &$<0.01$	\\
\hline \hline
\end{tabular}  
}
\end{table}

\subsection{ARIMA Modeling}

ARIMA is one of the most elementary as well as frequently used models for non-stationary time series. In astronomy, ARIMA has been applied to modelling radio emission of compact sources \citep{2001ApJS..136..265L} and exoplanet light curves \citep{2019AJ....158...57C}. The parameters $p$, $d$ and $q$ imply fitting a time series made stationary after differencing $d$ times to a combined autoregressive (AR) model of $p$ coefficients $\{ \phi_i\}$ and a moving average (MA) model of $q$ coefficients $\{ \theta_i\}$ with white noise terms $\epsilon_k$. Using the lag operator $L$, this can be expressed as,

\begin{equation}
\left( 1 - \sum_{i=1}^p \phi_i L^i \right)
(1-L)^d X_k
= \left( 1 + \sum_{i=1}^q \theta_i L^i \right) \epsilon_k \
\end{equation}

We choose the optimum model by minimizing the Akaike information criterion (AIC) \citep{2014ApJ...788...33K} for ranges of $p$ between $0$ and $10$ and $q$ between $0$ and $2$. Whereas as per the results of the ADF test, $d$ is kept fixed at $1$. The AIC \citep{akaike1973} accounts for both the quality of fit and the requirement that we use the fewest number of parameters to still be able to explain the data (in order to avoid over-fitting).

As displayed in Table \ref{Table:arima}, we find that for most of the light curves, the best-fit ARIMA are simple models with mostly $1$ AR and $2$ MA coefficients. The only exceptions are obs1-2000 and obs2-2001 of Mrk~421 which have long memories, i.e., $8$ and $5$ AR coefficients, respectively. These models are further validated by the application of Ljung-Box test \citep{10.1093/biomet/65.2.297} on the ACF of the residuals to test the null hypothesis that those are distributed as per a white noise process. In all of these cases the null hypothesis cannot be rejected because the minimum p-value is greater than $0.1$ for ACF time lags $< 50$ ($\sim 4800 s$). Hence, we find that the ARIMA model with the minimum AIC, as displayed in Table \ref{Table:arima}, are appropriate fit to the respective data.  

\begin{table}
\centering
{\footnotesize
  \caption{Best fit parameters for ARIMA models applied to the blazar light curves. \label{Table:arima}}
  \begin{tabular}{ccccccc} \hline\hline  
		Source &Observation &$p$ &$q$ &$d$ &$AIC$ &p-value$_{min}$\\
         & & & & & &  \\ \hline
		PKS &obs1-2004 &$2$ &$2$ &$1$ &$1162.35$ &$0.86$ \\
			2155-304&obs2-2007 &$1$ &$2$ &$1$ &$2231.68$ &$0.15$ \\
			&obs3-2010	&$1$ &$2$ &$1$ &$1772.24$ &$0.37$ \\
			
		3C~273 &obs1-2001 &$1$ &$2$ &$1$ &$2550.25$ &$0.73$  \\
		    &obs2-2007	&$1$ &$1$ &$1$ &$2186.24$ &$0.25$ \\
		    
		Mrk~421  &obs1-2000 &$8$ &$2$ &$1$ &$1490.15$ &$0.67$ \\ 
			&obs2-2001 &$5$ &$1$ &$1$ &$1814.81$ &$0.69$ \\
		   &obs3-2002 &$1$ &$2$ &$1$ &$988.23$	&$0.25$ \\
		   &obs4-2003 &$1$ &$2$ &$1$ &$1831.02$ &$0.31$ 	\\
\hline \hline
\end{tabular}  
}
\end{table}

\section{Frequency Domain Analysis} \label{sec:spectralanalysis}

For an observed light curve ${X_k}$ of duration $T$ sampled at $N$ instances ${t_k}$, the PSD $P(\nu)$ with the fractional rms squared normalization is defined as,
\begin{equation}
P(\nu_i)=\frac{2T}{\mu^2 N^2} \left [[\sum_{k=1}^{N} X_k cos(2\pi\nu_i t_k) ]^2 + [\sum_{k=1}^{N} X_k sin(2\pi\nu_i t_k) ]^2 \right]
\end{equation}
with $\nu_i = i/T$ , $i=1,...,N/2$, where the maximum frequency is the Nyquist frequency $\nu_{Nyq}=N/2T$ and $\mu$ is the mean. With this normalization, the PSD from different sources and/or measured by different instruments can be compared amongst themselves. 

There are limitations associated with the way PSD is calculated which include distortions due to finite length and irregular sampling of the light curves. These introduce error due to red noise leak and aliasing. This is alleviated to some extent in the Power Spectral Response (PSRESP) method \citep{1992ApJ...400..138D,2002MNRAS.332..231U,2008ApJ...689...79C}. Several light curves corresponding to a particular power spectrum shape are simulated using the algorithm by \citet{1995A&A...300..707T}. The sampling pattern of the observed light curve is applied followed by rebinning and interpolation, after which the PSD of those light curves are estimated. 

A pseudo-$\chi^2$ is calculated to further furnish a $F_{succ}$ which indicates the goodness-of-fit of the particular model in consideration. The power spectrum shape for which $F_{succ}$ is the greatest is considered to be the best estimate for the shape of the observed PSD. The uncertainty of the shape parameter(s), e.g., value of the power-law slope $\alpha$ is given by the slope for which $F_{succ}$ is 68.3\% of its maximum value.

In this study we simulate PSDs with bending power-law shapes, parameterized by a high-frequency slope $\alpha$, a low frequency slope $\beta$ and the break frequency $\nu_B$,
\begin{equation} \label{eq:psd_shape}
P(\nu)=A \nu^{-\beta} [ 1 + (\frac{\nu}{\nu_B})^{(\alpha - \beta)} ]^{-1} + P_{Poisson},
\end{equation}
Here we compute the normalised Poisson noise $P_{Poisson}$ as follows:
\begin{equation}
P_{Poisson}=\frac{\overline{\sigma_{err}^2}}{\mu^2 (\nu_{Nyq} - \nu_{min})}
\end{equation}

Based on the length and sampling rate of the \xmm{}  light curves, the range of frequencies we consider are between $~10^{-5.0}$ Hz and $~10^{-2.5}$ Hz. The red noise leak is minimized by extracting $200$ light curves from a very long light curve simulated using Equation \ref{eq:psd_shape}. In the case of Mrk~421 we fix the parameters $\beta = 1.2$ and $\nu_B = 10^{-6}$ Hz as given by \citet{2018ApJ...859L..21C}, and for 3C~273 and PKS~2155-304, we use $\beta = 1.0$ and $\nu_B = 10^{-6}$ Hz based on \citet{2008bves.confE..14M} and \citet{2001ApJ...560..659K}.
Generally, the high-frequency slope used to fit blazar PSDs is in the range $\alpha \gtrsim 2$ \citep[e.g.,][]{2010ApJ...719.1433A,2012ApJ...749..191C}. 
We apply the PSRESP method for a range of test slopes $1.0$ to $3.0$ in steps of $0.05$. We show the high-frequency slope which gives the greatest $F_{succ}$ along with its upper bounds in Table \ref{Table:psresp}. 
 
\begin{figure}
\includegraphics[scale=0.19,trim={1cm 8cm 0 10.5cm},clip]{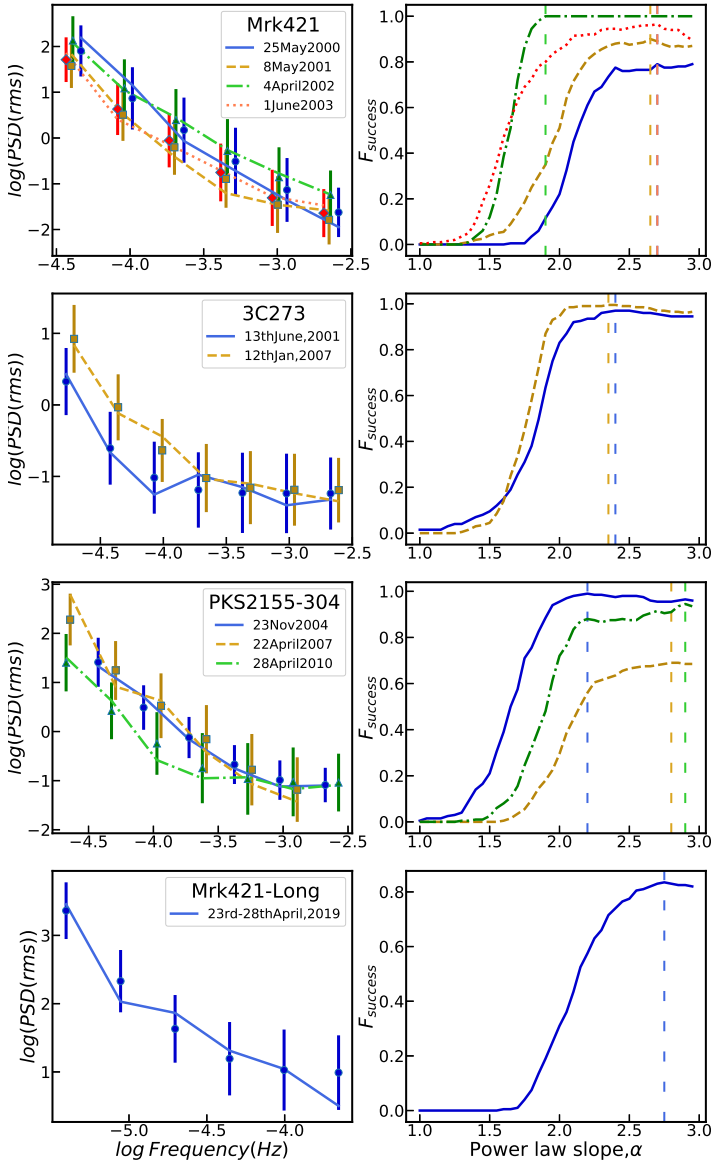}
\caption{The left panels depict the PSDs of the X-ray variability with \textit{solid/dashed/dot-dashed/dotted lines} corresponding to different epochs of observations for Mrk~421, 3C~273 and PKS~2155-304. The \textit{filled circles/squares/triangles/diamonds} with error bars are the mean and rms deviation for the best-fitting PSD shape as given by PSRESP. The right panels depict the respective success fraction distribution $F_{succ}$ for each light curve and their best-fitting high-frequency slope $\alpha$ for a bending power-law (\textit{spaced dashed lines}). \label{Figure:psresp}  }
\end{figure}

The best-fit power-law slope of the PSDs of Mrk~421 across the four epochs of observation are the same within the uncertainties (Figure \ref{Figure:psresp}, \textit{1st row}). The average flux during obs3-2002 is smaller than that during the rest of the epochs by a factor of $\sim 5$ but that does not affect the PSD shape or normalization during that epoch significantly. For 3C~273, we observe no significant change in the flux or the slope estimate of the PSD (Figure \ref{Figure:psresp}, \textit{2nd row}). The average flux of PKS~2155-304 during the second epoch (obs2-2007) is higher by a factor of $\sim 2$ compared to the other two epochs. However, it is the third observation from 2010, for which the PSD normalization is different (Figure \ref{Figure:psresp}, \textit{3rd row}).
Flattening of the PSDs is observed at frequencies $\gtrsim$ 1 mHz due to Poisson noise or noise introduced due to the sampling pattern.

We also look at the nature of the PSD for the sub-week length light curves of Mrk~421. Since the shapes of the PSDs corresponding to the 1998 \asca{} and 2017 \astst{} observations have already been studied in \citet{2001ApJ...560..659K} and \citet{2018ApJ...859L..21C}, respectively, here we conduct a PSRESP analysis of the \astst{} data from 2019. Proceeding with the same method that we use for \xmm{} light curves, we find a best fitting simple power-law slope of $2.75^{+\infty}_{-0.55}$ with a success fraction of $0.84$ (Figure \ref{Figure:psresp}, \textit{4th row}). The value of the power-law slope is consistent with the high-frequency slope that we estimated earlier. 

\begin{table}
\centering
{\footnotesize
  \caption{Results of applying PSRESP to the blazar light curves. \label{Table:psresp}}
  \begin{tabular}{ccccc} \hline\hline  
		Source &Observation &Slope &Success &$P_{Poisson}$ \\ 
         & &$\alpha$ & Fraction &   \\ 
         &  &  & $F_{succ}$ &  \\ \hline \\

		PKS&obs1-2004 &$2.20^{+\infty}_{-0.46}$   &$0.99$ &$0.788$	 \\
			2155-304 &obs2-2007	&$2.80^{+\infty}_{-0.68}$  &$0.69$	&$0.409$ \\
			&obs3-2010	&$2.90^{+\infty}_{-0.93}$  &$0.95$	&$0.939$ \\
		
		3C~273 &obs1-2001	&$2.40^{+\infty}_{-0.49}$  &$0.97$ &$0.583$  	\\ 
		    &obs2-2007	&$2.35^{+\infty}_{-0.52}$  &$0.99$  &$0.649$\\
		
		Mrk~421  &obs1-2000 &$2.70^{+\infty}_{-0.55}$ &$0.79$ &$0.171$ \\ 
			&obs2-2001 &$2.65^{+\infty}_{-0.61}$ &$0.9$ &$0.127$ \\
		   &obs3-2002 &$1.90^{+\infty}_{-0.21}$ &$1.0$ &$0.430$	\\
		   &obs4-2003 &$2.70^{+\infty}_{-0.97}$ &$0.97$ &$0.169$	\\
\hline \hline
\end{tabular}  
}
\end{table}

\section{Variance Analysis} \label{sec:variance}
The variance $S^2$ of a light curve is linked with the normalized PSD through the following relation:
\begin{equation}
S^2=\sum_{j=1}^{N/2} P(\nu_j) \Delta \nu
\end{equation}
Therefore, in addition to comparing the PSD shapes from various epochs of observation, studying the nature of variability is possible by looking at the statistical properties of variance. 

\subsection{Scatter of Excess Variance} \label{sec:scatterofvar}

The distribution of the variance estimates depends on the PSD power-law slope of the underlying stochastic process \citep{2003MNRAS.345.1271V}. Whereas this distribution for a stationary white-noise process ($P(\nu) \sim const.$) is a Gaussian, for a globally non-stationary red noise process ($P(\nu) \sim \nu^{-2}$) the distribution becomes significantly broad and skewed. Therefore, fluctuations in variance larger than that expected to arise from the power-law nature of the PSD implies the presence of local stationarity. This guides the usage of the method of scatter of excess variance. We break up the light curve into segments of 50 data points and calculate the sample variance $S^2$ for each segment. In order to take into account the measurement errors in the counts $\sigma_{err,i}$, we obtain the excess variance by substracting from the sample variance the mean squared error, 
\begin{equation}
\sigma_{XS}^2 = S^2 - \overline{\sigma_{err}^2}
\end{equation}

Stationarity is determined by how well bounded the scatter of $\sigma_{XS}^2$ is within the $90 \%$ and $99 \%$ intervals computed in \citet{2003MNRAS.345.1271V}.
Drawing from the results of section \ref{sec:spectralanalysis} where all the blazars have high-frequency power-law slopes $\sim$2.5, we find it suitable to use the expected bounds on the scatter computed for a process with a power-law of slope $2.5$. These intervals turn out to be ($-0.95,+0.51$) and ($-1.31,+0.82$) about the logarithm of the mean $\sigma_{XS}^2$ respectively. In some segments, negative values of $\sigma_{XS}^2$ arise because the variance due to measurement errors dominates over that due to the red-noise nature of the variation. In these cases, we reject the variance estimates. 

We calculate the uncertainties of the excess variance estimates using the variance of the sample variance estimate \citep{CaseBerg:01}, and subtract from it the contribution from the measurement errors. Through error propagation it can be shown that:
\begin{equation} \label{eq:exvarerr}
err(log(\sigma_{XS}^2))=\frac{1}{{\sigma_{XS}^2}} \sqrt{\frac{2}{N-1} S^4 - \frac{Var(\sigma_{err}^2)}{N}}
\end{equation}
This results in larger uncertainties for smaller values of $\sigma_{XS}^2$, as can be seen in the scatter of variances plot for 3C~273.

\begin{figure*}
\begin{center}
\includegraphics[scale=0.79,trim={0.5cm 7cm 0.6cm 7.5cm},clip]{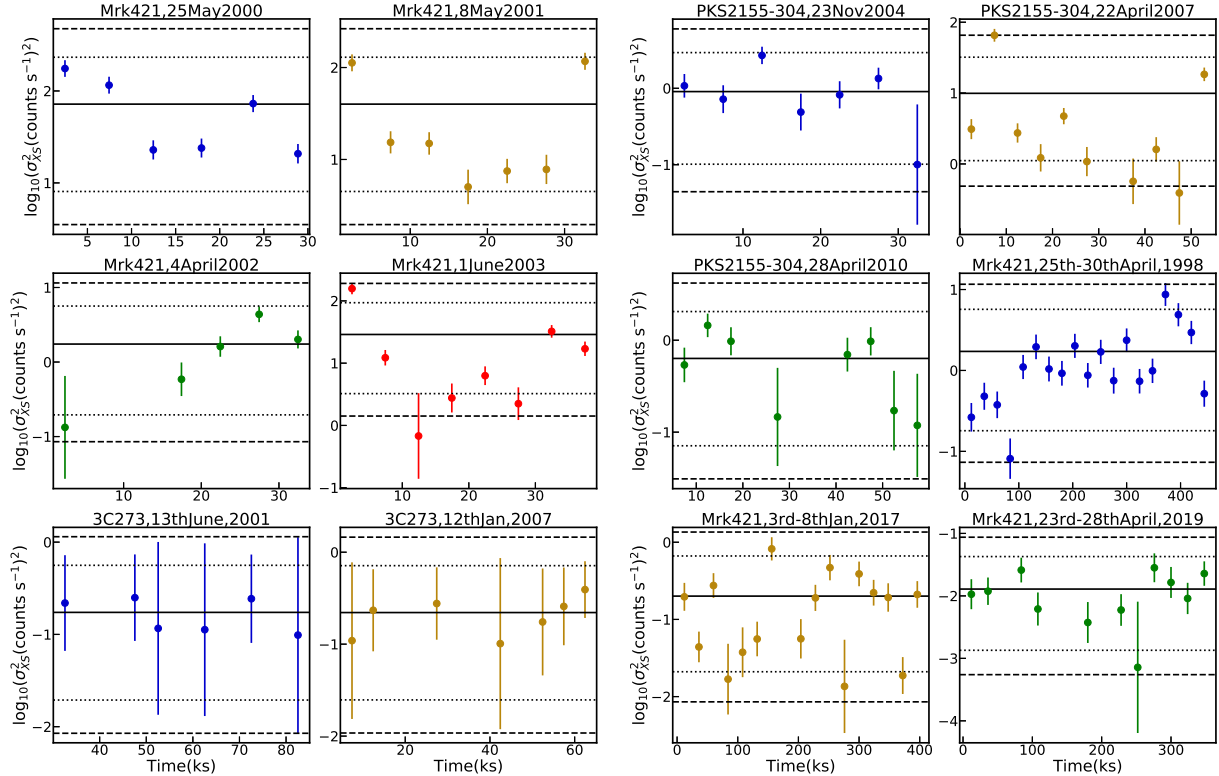}
\caption{Scatter in log($\sigma_{XS}^2$) along with their mean (\textit{solid line}), 90\% interval (\textit{dotted line}) and 99\% interval (\textit{dashed line}) for the sub-day light curves of  Mrk~421, 3C~273, PKS~2155-304 and the sub-week light curves of Mrk~421. The error bars have been calculated using equation \ref{eq:exvarerr}. \label{Figure:exvar} }   
\end{center}
\end{figure*}

We observe that for Mrk~421 the scatter is well within the 90\% interval of the expected scatter except for obs4-2003 (Figure \ref{Figure:exvar}). When the bounds pertaining to a power-law slope of $2.0$ is used, the data points are no longer within the $90 \%$ interval. We note that the mean variance of obs3-2002 is an order of magnitude less than that of the other observations. In 3C~273, the scatter is properly bound given a PSD slope of $2.5$ (Figure \ref{Figure:exvar}). Besides, the mean of the variance estimates are almost the same for obs1-2001 and obs2-2007. In the case of PKS~2155-304, we observe that the excess variance in obs1-2004 and obs3-2010 are well within the expected scatter assuming a PSD slope of $2.5$ while that for obs2-2007, in which the average flux is larger by a factor $\sim$2, are not (Figure \ref{Figure:exvar}). The mean value of variance for this observation is higher by an order of magnitude.

Considering the sub-week length light curves of Mrk~421, we first perform rebinning with a common sampling interval of 1200s.
We find that the scatters of $\sigma_{XS}^2$ slightly deviate from the bounds set by a PSD slope of $2.5$ (Figure \ref{Figure:exvar}). We also note that the mean value of the variance according to the 2019 observation is very small in magnitude compared to the other two observations. 

\subsection{Fractional Variance} \label{sec:fractionalvar}
In this subsection, we calculate the fractional variance of the entire light curves and compare its value from one epoch to another. That is a widely applied test of stationarity \citep{2005ApJ...629..686Z, 2019MNRAS.482.2088A}. The fractional rms variability is calculated as the square root of the excess variance normalized with respect to the mean flux of that segment:
\begin{equation}
F_{var} = \sqrt{\frac{\sigma_{XS}^2}{\bar{X}^2}}
\end{equation}


We perform a crude estimation by calculating the mean $F_{var}$ along with its standard error for an entire light curve and comparing such values among two or more light curves of the same source. The estimates of excess variance $\sigma_{XS}^2$ that we calculate for every 50 data points in the previous section are used to compute a value of $F_{var}$. As earlier, we reject the negative values of $\sigma_{XS}^2$. Approximately, 5-10 estimates of fractional variance are obtained for each light curve, from which we find the average $F_{var}$ and its standard error, as displayed in Table \ref{Table:fvar}. 

We find that obs1-2000 of Mrk~421 has a significantly larger value ($\sim3\sigma$) of $F_{var}$ compared to the later three observations. Even though there is a sharp change in flux between the second and third observations, it is not reflected in the corresponding $F_{var}$ estimates. In 3C~273, no significant change in the fractional variance is observed. However, in the observations of PKS~2155-304, we notice a significant change in the fractional variance for the observation during obs2-2007 which may be linked to the large changes in the mean flux and scatter of variances noticed in the previous analyses for the same epoch of observation.

We find that for the $\sim$400 ks long observations of Mrk~421,  the $F_{var}$ estimated in the aforesaid fashion is larger compared to the ones from \xmm{} light curves with sampling intervals of $96$ s. This is natural considering the red noise nature of the light curves, in which the amount of variability, here quantified by $F_{var}$, is greater at longer timescales. Nonetheless, we see that the $F_{var}$ for the 1998 \asca{} light curve is almost twice that of the $F_{var}$ values from the 2017 and 2019 \astst{} data.

\begin{table}
{\footnotesize
\centering
  \caption{Average fractional variances of the blazar light curves
   \label{Table:fvar}}
  \begin{tabular}{cccc} \hline\hline 
                Source  &Observation   &$N_{seg}$ &$F_{var}$       \\ 
                 \hline \\
                
                PKS  &obs1-2004 &7 &$1.86 \pm 0.31 $  \\
                2155-304 &obs2-2007 &10   &$2.25 \pm 0.66 $  \\
                         &obs3-2010 &8   &$1.71 \pm 0.27 $   \\
                              
                3C~273  &obs1-2001  &6   &$0.60 \pm 0.05 $  \\
                             &obs2-2007  &7   &$0.76 \pm 0.07 $  \\

                Mrk~421  &obs1-2000 &6  &$2.83 \pm 0.44$\\ 
                         &obs2-2001  &7  &$1.53 \pm 0.43$\\
                         &obs3-2002 &5  &$1.68 \pm 0.45$ \\
                         &obs4-2003 &8  &$1.45 \pm 0.45$ \\
                \hline \\
                Mrk~421  &1998 &7  &$9.42 \pm 1.50$\\ 
                         &2017  &5  &$3.69 \pm 1.32$\\
                         &2019 &6  &$4.75 \pm 0.64$ \\
\hline \hline
\end{tabular}  
}
\end{table}

\newpage
\section{Flux-rms Relationship} \label{sec:linearfluxrms}

\begin{figure}
\includegraphics[scale=0.72,trim={0.5cm 0.3cm 8cm 0.3cm},clip]{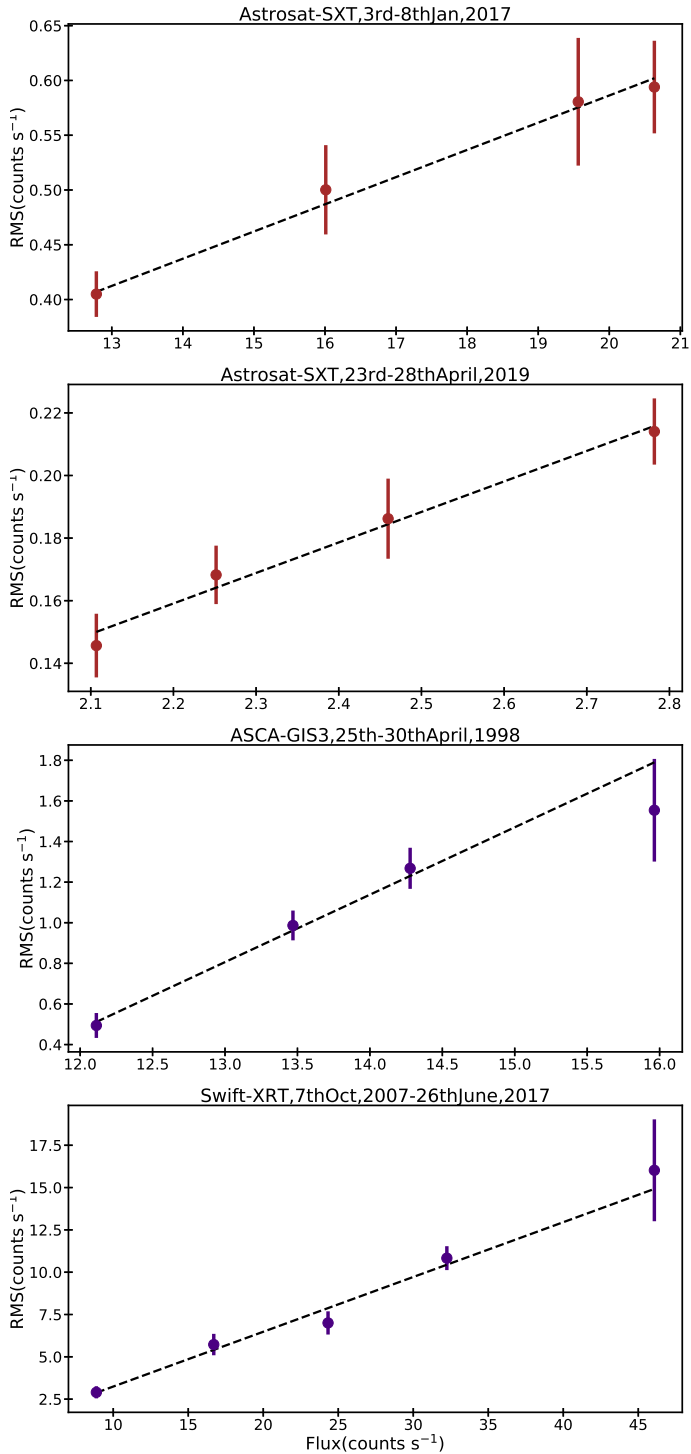}
\caption{Linear flux-rms relations from \astst{} SXT, \asca{} GIS3 and \swift{} XRT observations of Mrk~421. The \textit{indigo} and \textit{brown filled circles} correspond to the rms value in each flux bin calculated with and without subtracting measurement errors, respectively. Their error bars are the scatter of rms in each flux bin. The \textit{black dashed line} is the best-fitting straight line for these points \label{Figure:fxrms}}
\end{figure}

We investigate the presence of flux-rms relationship \citep{2005MNRAS.359..345U} in the $\sim$400 ks X-ray light curves of Mrk~421 from \astst{} SXT and \asca{} GIS3. For the \astst{} SXT observation in the $0.3-7.0$ keV energy band made during 2017 January 3-8, we take segments of the light curve of length 11 ks and arrange those in ascending order of mean flux. Alongside, we compute the rms for each segment and then bin together 9 such segments to generate a mean flux and rms for each bin. The standard error is calculated from the scatter in rms estimates in each bin. We plot the results of 4 such bins and observe a distinct linear correlation with a best fit slope of 0.025$\pm$0.001 counts s$^{-1}$ (Figure \ref{Figure:fxrms}, \textit{1st row}). Even when we take segments of 12.5 ks and 13.5 ks, the linear trend persists with minor change in the best fit slope estimates- 0.032$\pm$0.005 counts s$^{-1}$ and 0.025$\pm$0.002 counts s$^{-1}$ respectively.


For the second \astst{} SXT observation made during 2019 April 23-28 in the $0.3-7.0$ keV energy band, we use the above formulation but with segments of 10.5 ks, giving a linear flux-rms relation with a best-fit slope estimate of 0.097$\pm$0.009 cts\,s$^{-1}$ (Figure \ref{Figure:fxrms}, \textit{2nd row}), which is steeper by a factor of a few compared to the previous light curve. We note that the mean flux of this epoch is an order of magnitude smaller compared to the previous epoch. When we take segments of 13.5 ks and 12.5 ks, we get similar best-fit slope estimates of 0.106$\pm$0.002 cts\,s$^{-1}$ and 0.083$\pm$ 0.007 cts\,s$^{-1}$, respectively.

Similarly, we get a linearly correlated flux-rms relation using segments of 12 ks from the observation by \asca{} GIS3 in the $0.1-10$ keV band during 1998 April 25-30. However, this is accompanied by a steeper estimate of the slope of 0.329$\pm$0.044 . These data are well-sampled enough for us to calculate the rms after subtracting measurement errors \citep{2004MNRAS.348..783M}. Even after taking into account this effect, we observe that the flux-rms relation remains persistent, presenting a best-fit slope of 0.332$\pm$0.033 (Figure \ref{Figure:fxrms}, \textit{3rd row}).

We find a linear flux-rms relation in the long-term ($\sim$ 10 yr) X-ray observation of Mrk~421 carried out by \swift{} XRT. We use segments of 5 Ms ($\sim$ 60 days) and make 5 flux bins such that there are 12 segments in each bin. We find a best fitting slope of 0.324 $\pm$ 0.026 (Figure \ref{Figure:fxrms}, \textit{4th row}). This result is consistent with that of \citet{2016A&A...591A..83S}. In our case, the scatter in the flux-rms relation may be smaller because of the use of the aforesaid method. 

In the \astst{} light curves from 2017 and 2019, the best fit slopes ($\lesssim$0.1) are inconsistent with the slopes we obtain ($\sim$0.33) for the \asca{} observation in 1998 and the \swift{} observation between 2007-2017. This altered value of the slope may arise due to the measurement errors not being subtracted in the case of the flux-rms relation for the \astst{} light curves. 

In {\S}\ref{sec:fractionalvar}, we showed how the average level of $\sigma_{XS}^2$ tracked the average level of flux, noticeable especially when there is a large increase (obs3-2002 for Mrk~421) or decrease (obs2-2007 for PKS~2155-304) in flux. This might be the manifestation of an underlying flux-rms relation as has been noticed in both these sources before \citep{Chevalier:2015gia,2016A&A...591A..83S}.

\section{Discussion} \label{sec:disucss}

After having diagnosed the global non-stationarity of these light curves using the ADF test and with it justified the usage of ARIMA models, we find that most of the best fits to the observations correspond to short memory or low order processes with $p \leq 2$ and $q \leq 2$. This can be attributed both to the versatility of low order ARIMA models and also to the similarity of red noise emission from blazars \citep{2001ApJS..136..265L}. Notable exceptions are the light curves of Mrk~421, in which long memory processes are noticed for obs1-2000 and obs2-2001. A more detailed study with ARIMA modeling of similar blazar light curves possibly at multiple wave bands is necessary to test if such a change is linked to local non-stationarity, and in turn, to differences in the emission mechanism in the jet.

The results of the PSRESP method indicate that the best-fit power-law slopes for the PSDs of Mrk~421, 3C~273, and PKS~2155-304 are consistent across various epochs of observation within their levels of uncertainties. 
In the analysis of the scatter of $\sigma_{XS}^2$, we find that all the sources have reasonable bounds on their scatter by the $90 \%$ interval corresponding to a slope of $\alpha = 2.5$. Only in obs4-2003 of Mrk~421 and obs2-2007 of PKS~2155-304 do the scatters marginally exceed these bounds. Considering the estimates of $F_{var}$, obs1-2000 of Mrk~421 and obs2-2007 of PKS~2155-304 show deviations from the values of other observations of the respective source. Among the longer light curves of Mrk~421, particularly the \asca{} observation from 1998 April stands out- a scatter of variance that might exceed the expectation for a PSD with $\alpha = 2.5$ alongwith a relatively high $F_{var}$ may indicate local non-stationarity. 

In summary, based on the PSD only, all the sources may be qualified as not exhibiting local non-stationarity but there are minor evidences to the contrary when we consider the complementary analyses. While for PKS~2155-304 and 3C~273 we get signs of local non-stationarity which are not consistent among the various tests, for Mrk~421 we get multiple clues which may indicate local non-stationarity synchronous with large changes in its average flux.

While we do not find any significant change in the PSD between multiple epochs for the same blazar we must note that a small temporal change in the PSD of red noise variability may not always be detectable with the quality of data that is available.
There may be an excess scatter in the binned PSD estimates obtained from light curves spanning $\sim$hr to $\sim$yr leading to poor fits of a model PSD and subsequently higher rejection probabilities. Sampling window and Poisson noise give rise to distortions in the PSD, which are difficult to quantify, and cause uncertainties in the estimates of PSD slopes, which can eclipse any change in the slope of the PSD of the underlying process.

In alternative methods of PSD modelling, such as, CARMA, the difficulties in discerning a local non-stationary process persists. \citet{2014ApJ...788...33K} showed, using an irregularly sampled simulated light curve stitched together from two separate stationary CARMA processes with differing variances, that a CARMA (5,2) model provided a good fit as if the complete light curve was a single stationary process. It is as if the PSD of the combined process is a blend of the separate processes weighted according to their respective SNRs. 

The local non-stationarity in the PSD of the X-ray variability of Seyfert galaxies such as IRAS 13224-3809 has been explained as due to accretion rate close to the Eddington limit, or deformations in a time evolving corona \citep{2020NatAs.tmp....2A,2015MNRAS.449..129W}. In the blazars the X-ray emission is mostly from the jet and hence if it exhibits local non-stationarity that will imply some important change in the emission process inside the jet. Alternatively, the jet variability may be driven by fluctuations in the accretion disk and in that case the non-stationarity may be due to certain specific kind of changes in the disk. 

We find a linear flux-rms relation in the X-ray light curves of Mrk~421 separated by a period as long as two decades and over timescales of $\sim$days to $\sim$months. It is one of the few cases, in which flux-rms relation has been detected in a blazar across such long timescales. This, along with similar demonstrations in other blazars and other wave bands \citep{2013ApJ...766...16E, 2017ApJ...849..138K}, implies that the flux-rms relation may be a ubiquitous feature of blazar variability similar to other accreting black hole systems, e.g., BHXRBs. 

There is a significant difference in the best-fit slope of the flux-rms relation among the short term as well as the long term observation of Mrk~421 across 2 decades. This may be the result of some systematic effect like error subtraction or it may be due to local non-stationarity in the source itself. For BHXRBs the slope and the flux intercept are known to depend on the spectral state of the BHXRB \citep{2012MNRAS.422.2620H} and consequently on the variable shape of the PSD. In such cases the long term flux-rms relation cannot be held representative of its short term counterpart. 

Particularly, the fact that we find a flux-rms relation from the short-term \astst{} as well as the long-term \swift{} light curves may imply that the jet variability is not due to a simple scaling relation in which short-term variations are modulated by a single slower process \citep{2017A&A...601L...1U}. 
The so called ``minijets-in-a-jet'' is an additive type model \citep{2012sf2a.conf..567B} of randomly oriented minijets in the rest frame of the jet. The flux from each minijet is shown to have a Pareto distribution, and the resultant sum of such randomly placed minijets can reproduce the flux-rms relation as observed here, as well as the skewed non-Gaussian flux distribution. An alternative explanation of the log-normal flux distribution is Gaussian distributed perturbations in the electron acceleration or escape timescales by the shock front moving through the jet \citep{2018MNRAS.480L.116S}. 
More generally, the flux-rms relation implies that the variability process is multiplicative and that is a stringent constraint in the jet emission models. We note, however, that the direct equivalence between flux-rms relation, multiplicative nature and log-normal flux distribution as discussed by \citet{2005MNRAS.359..345U} has been disputed by \citet{2020arXiv200108314S}. According to the latter, the flux-rms relation may be a generic property of quasar variability and similar  time series and may not imply a specific type of emission and dynamics model as discussed above. 

\citet{2018MNRAS.480.2054M} show a connection between multi-wavelength correlation properties of jet variability with the power spectrum of the bulk Lorentz factor fluctuations caused by variations in the disk. However, \citet{2019MNRAS.486.1672M} have shown that a direct coupling of a characteristic timescale in the disk and that in the jet variability does not occur. Therefore, local stationarity or a lack thereof, or the presence of a flux-rms relation in the jet emission variability may place strong constraints on the mode of disk-jet connection and the variability mechanism in the jet itself, e.g., dynamics of shocks propagating down the jet, radiation cooling, and turbulence such that models like the above may be tested.  

\section{Acknowledgements}
We thank the anonymous referees whose suggestions helped us improve this manuscript. We thank P. Uttley for a very useful discussion regarding the applicability of time and frequency-domain methods in the analysis of AGN variability. We thank R. Misra whose question prompted the beginning of this investigation. SB acknowledges the DST INSPIRE and JBNSTS fellowship. RC thanks Presidency University for support under the Faculty Research and Professional Development (FRPDF) Grant, ISRO for support under the Usage of \astst{} Archival Data Grant, and IUCAA for their hospitality and usage of their facilities during his stay at different times as part of the university associateship program. RC received support from the UGC start-up grant. This work has made use of data from the \astst{} mission of the ISRO, archived at the Indian Space Science Data Centre (ISSDC), and that from \xmm{}, software and/or web tools obtained from NASA's High Energy Astrophysics Science Archive Research Center (HEASARC), a service of Goddard Space Flight Center and the Smithsonian Astrophysical Observatory. The \textit{statsmodel.tsa} \citep{seabold2010statsmodels} module of Python is used for the respective time series analysis. 

\bibliographystyle{aasjournal} \bibliography{mybib}

\end{document}